\documentclass[times]{nlaauth}
\usepackage{moreverb}
\usepackage{amsmath}
\usepackage[noend]{algorithmic}
\usepackage{lipsum}

\begin{document}


\title{Fast low-rank approximations of multidimensional integrals in ion-atomic collisions modelling}

\author{M.~S.~Litsarev\affil{1}\corrauth, I.~V.~Oseledets\affil{1,2}}
\address{
\affilnum{1}Skolkovo Institute of Science and Technology,
Novaya St. 100, Skolkovo, 143025  Moscow Region, Russia\break 
\affilnum{2}Institute of Numerical Mathematics, Gubkina St. 8, 119333 Moscow, Russia}
\corraddr{m.litsarev@skolkovotech.ru}

\begin{abstract}
An efficient technique based on low-rank separated
approximations is proposed for computation
of three-dimensional integrals arising in the energy deposition model that describes ion-atomic collisions.
Direct tensor-product quadrature requires grids of 
size $4000^3$ which is unacceptable. 
Moreover, several of such integrals have to be computed
simultaneously for different values of parameters.
To reduce the complexity, we use the structure of the 
integrand and apply numerical linear algebra techniques for the
construction of low-rank approximation. 
The resulting algorithm is $10^3$ faster than spectral quadratures in spherical 
coordinates used in the original DEPOSIT code. 
The approach can be generalized to other multidimensional problems in physics.
\end{abstract}


\maketitle

\vspace{-6pt}

\section{Introduction}
\vspace{-2pt}

Computation of multidimensional integrals is one
of the most time-consuming tasks in physics. 
Standard approaches either have high complexity or
 require sophisticated quadrature schemes. 
Already in a three-dimensional case the integrand 
may depend on many parameters and should be 
computed many times, so the computational cost of the 
simplest tensor-product quadratures is an important issue.

One of the promising tools to reduce the dimensionality of the problem
(and hence the number of mesh points where the integrand should be calculated)
is the usage of \emph{separation of variables} in the integrand
(see, for example, \cite{beylkin-2002,vkhs-2el-2013}).
The idea was known for a long time 
(for examples we refer the reader to the review~\cite{kolda-review-2009}),
but it has become practically useful after the fast algorithms 
of decompositions of functions in a  \emph{separated form}
had been obtained in two-~\cite{tee-cross-2000,bebe-2000}, 
three-~\cite{ost-tucker-2008, crossconv2015} 
and multidimensional cases~\cite{ot-tt-2009, osel-tt-2011}.

Let $F(x, y)$ be a function of two variables $x, y$ where point $(x, y)$ is
in a certain rectangle $[a_x,b_x] \otimes [a_y,b_y]$. 
The function is said to be in a \emph{separated form}
if it can be represented as a sum of products of univariate functions:
\begin{equation}
\label{FxyCanonical}
F(x,y)=\sum_{\tau=1}^{q} \sigma_{\!\tau} \,u_{\tau}(x)g_{\tau}(y).
\end{equation}
The minimal number $q$ such that \eqref{FxyCanonical} exists 
is called \textit{separation rank}. 
Direct generalization of \eqref{FxyCanonical} to multivariate functions
is referred to as a canonical polyadic (CP, also known as
CANDECOMP/PARAFAC) decomposition~\cite{kolda-review-2009,parafac1970}. 
The reader can find examples of applications
of separated representations in multidimensional cases
in~\cite{beylkin-2002, osel-constr-2013, beylkin-high-2005,
khor-low-rank-kron-P1-2006, khor-low-rank-kron-P2-2006,
khor-acc-2010, khor-qtt-2011,khor-prec-2009, khor-ml-2009,mpi-chem3d-2009}.

For a function given in the separated form
the integration is simplified a lot. Indeed,
\begin{equation}
\label{lrint:2dex}
\iint F(x,y) dx dy = \sum_{\tau=1}^{q} \sigma_{\tau} 
\! \int_{a_x}^{b_x} \!\!\! u_{\tau}(x) dx \! \int_{a_y}^{b_y} \!\!\! g_{\tau}(y) dy,
\end{equation}
and the problem is reduced to the computation of one-dimensional integrals,
which can be computed using fewer quadrature points than the original integral.

In case of a discrete approximation of \eqref{FxyCanonical}   
\begin{equation}
\label{lrint:2ddiscr}
F(x_i, y_j) \approx 
\sum_{\tau=1}^{q} \sigma_{\tau} u_{\tau}(x_i) g_{\tau}(y_j)
\end{equation}
with the error $\varepsilon$ in the Frobenious norm
the number $q$ is called $\varepsilon$-rank. 
We shall assume the notion of $\varepsilon$-rank 
when the term \textit{rank} will be mentioned in the text bellow.
Expression~\eqref{lrint:2ddiscr} can be rewritten in the matrix form
\begin{equation}
\label{AapproxUSigVt}
A \approx U \Sigma G^{\top},
\end{equation}
where $A$ is an $n \times m$ matrix with elements $A_{ij}=F(x_i, y_j)$, 
$U$ is an $n \times q$ matrix with elements $U_{i\tau}=u_{\tau}(x_i)$, 
$G$ is an $m \times q$ matrix with elements $G_{\! j\tau}=g_{\tau}(y_j)$
and $\Sigma$ is a $q \times q$ diagonal matrix with elements $\sigma_{\tau}$ on the diagonal. This is a standard \emph{low-rank approximation problem} for a given matrix.
Provided that a good low-rank approximation exists, there are very efficient \emph{cross approximation algorithms} \cite{tee-cross-2000,bebe-2000} that need only $\mathcal{O}((n + m)q)$ elements of a matrix to be computed. 

In this paper we describe how to apply this technique to speedup 
computations of three-dimensional integrals
in the energy deposition model intended to describe
ion-atomic collisions.
This model was introduced by N.~Bohr~\cite{bohr1915} and 
developed further by
A.~Russek and J.~Meli~\cite{RussekMeli1970}, 
C.L.~Cocke~\cite{cocke_pra1979},
and V.P.~Shevelko~\textit{at al.}~\cite{litsarev-jpb-2008}.
Theoretical development of the model is presented 
in~\cite{litsarev-jpb-2008, litsarev-jpb-2010, litsarev-nimb-2009, litsarev-jpb-2009}.
Examples of calculations are reported
in~\cite{litsarev-springer-2012, litsarev-hci-2012, uspekhi2013, litsarev-jpb-2014}.
Detailed description of the computer code DEPOSIT 
and user guide are given in~\cite{litsarev-cpc-2013}
and its updated version based on the separated representations 
is avialable in~\cite{litsarev-cpc-2014}.

The code DEPOSIT allows to calculate total and multiple electron loss
cross sections and ionization probabilities needed for estimation of losses and lifetimes of fast ion beams, background pressures and pumping requirements in accelerators and storage rings.
All of them are, in fact, functionals of the deposited energy~$T(b)$
($b$ is the impact parameter of the projectile ion), 
which in turn is a three-dimensional integral over the coordinate
space. To calculate any of these parameters one has to compute
$T(b)$ in all points of the $b$-mesh.

In the original work~\cite{litsarev-cpc-2013} an advanced 
quadrature technique was used, and the computational time
has appeared to be much faster in a comparison 
with direct usage of uniform meshes. 
Calculation of the deposited energy $T(b)$ for  
a given atomic shell in one point~$b$ took about several seconds.
For complex ions it was about few 
hours on one processor core in total, that is not enough fast. 
To overcome this issue a fully scalable parallel variant of the
algorithm was proposed, but the computational time was still large.

We present an entirely different approach for the computation of
$T(b)$ in many points of the $b$-mesh based on 
the idea of separation of variables~\eqref{FxyCanonical}.
An approximation of functions to be integrated by a sum of products of
univariate functions allows to effectively decrease the dimensionality of the
problem. This requires active usage of numerical and analytical tools. 

The problem setting is as follows. 
The deposited energy $T(b)$ is a three-dimensional integral  
\begin{equation}
\label{TbDef}
T(b)=\iiint \! \Delta E(x,z - b) \rho(x,y,z) dx dy dz.
\end{equation}
It involves cylindrically symmetric function of two variables
(the energy gain $\Delta E$ during an ion-atomic collision)
which is a smooth finite function  and 
spherically symmetric function of three variables 
(electron density in a Slater-type approximation)
which decays exponentially.  
For details and definitions we refer the reader to Appendix~\ref{App:PhysModel}.
Previous approach \cite{litsarev-cpc-2013} used tensor-product 
quadratures in spherical coordinates. 
We use very fine uniform meshes and low-rank approximation.

In Section~\ref{RhoExpSect} the Slater-type 
function of three variables is decomposed by 
the exponential sums approach~\cite{beylkin-expsum-2005,beylkin-expsumrev-2010}. 
The integral is immediately reduced to a two-dimensional one of a
simpler structure. In Section~\ref{FastTb2DSect}
for a function of two variables we use the pseudo-skeleton
decomposition of matrices~\cite{tee-mosaic-1996,
gtz-psa-1997, gtz-maxvol-1997}
computed via a variant of the incomplete cross
approximation algorithm~\cite{tee-cross-2000}. 
Combining these two representations we obtain then
an efficient algorithm with $\mathcal{O}(n)$ scaling, 
in comparison with $\mathcal{O}(n^3)$ complexity for direct integration 
over a three-dimensional mesh.
We show numerically that the function in question 
can be well-approximated by a separable
function. 
Thus, the approximation can be computed in $\mathcal{O}(n)$
time, where $n$ is the number of grid points in one dimension.
The computation  of $T(b)$ on the whole $b$-mesh takes less then one
minute instead of several hours and total speedup of the program is about $\sim 10^3$ times.
Illustrative examples are given in Section~\ref{NumericSect}.
All the equations related to the physical model
are written in atomic units.

\vspace{-6pt}

\section{Numerical procedure}
\label{sectTb}

\subsection{Exponential sums.}
\label{RhoExpSect}

In this section we present analytical expansion of the spherically symmetric
electron density $\rho_{\gamma}(r)$ in Cartesian coordinates 
as a sum of separable functions.
We use this decomposition later in Section~\ref{FastTb2DSect} 
for analytical integration in one dimension.

A three-dimensional electron density $\rho_{\gamma}(r)$ is taken 
in a Slater-type approximation
\begin{equation}
\label{rhoDef}
\rho_{\gamma}(r)=C_{\gamma} r^{\alpha_{\gamma}} e^{-2\beta_\gamma r} ,
\qquad r=\sqrt{x^2 + y^2 + z^2},
\end{equation}
where integer parameter $\alpha_{\gamma}$ and
real $C_{\gamma}$  and $\beta_\gamma$
correspond to one atomic shell labeled by~$\gamma$
(see also Appendix~\ref{App:Laplace} for description of  parameters).
For the density $\rho_\gamma(r)$ the following normalization condition occurs
\begin{equation}
\int_{0}^{\infty} \rho_{\gamma}(r) dr=N_{\gamma},
\end{equation}
where $N_{\gamma}$ is the number of electrons 
in a $\gamma$-shell.
For the sake of simplicity index~$\gamma$ will be skipped
and only one shell will be considered in equations bellow.

For a function $\rho(x, y, z)$ defined in \eqref{rhoDef} 
the separation of variables~\eqref{FxyCanonical} in Cartesian coordinates
can be done analytically~\cite{beylkin-expsum-2005, beylkin-expsumrev-2010,
hackbra-expsum-2005, GHK-ten_inverse_ellipt-2005}.  
The main idea is to approximate the Slater density function by a sum of Gaussians
\begin{equation}
\label{gaussianExpand}
\rho(r) \approx \sum_{k=0}^{K} \lambda_k e^{-\eta_k r^2}.
\end{equation}
Once the approximation \eqref{gaussianExpand} is computed, the
separation of variables in Cartesian coordinates is immediately done
\begin{equation}
\label{rhoSeparated}
\rho(x,y,z) \approx \sum_{k=0}^{K} \lambda_k \,
e^{-\eta_k x^2} \, e^{-\eta_k y^2} e^{-\eta_k z^2}.
\end{equation}

The technique for computation of nodes $\lambda_k$ and weights $\eta_k$ is
based on the computation of inverse Laplace transform.
Let us consider a function $f_{\alpha \beta}(t)$ such that its Laplace
transform is a function $F_{\alpha \beta}(s)$:
\begin{equation}
\label{FabsqS}
F_{\alpha \beta}(s)=
\frac{\rho(\!\sqrt{s}\,)}{C}  
={\left(\!\sqrt{s}\right)}^{\alpha} e^{-2\beta \!\sqrt{s}}=
\int^{\infty}_{0} e^{-st}   f_{\alpha \beta}(t) \,dt,
\end{equation}
where $\alpha$ and $\beta$ are parameters
of Slater density~\eqref{rhoDef}.
Inverse Laplace transform $f_{\alpha \beta}(x)$ can be 
computed analytically for known $F_{\alpha \beta}(s)$~\eqref{FabsqS}. 
In Appendix~\ref{App:Laplace} we present explicit expressions 
for functions $f_{\alpha \beta}(t)$ corresponding to
function~\eqref{FabsqS} for integer $\alpha$ and real positive $\beta$.

Once function $f_{\alpha \beta}(t)$ 
in expression~\eqref{FabsqS} is known, the 
integral~\eqref{FabsqS} can be computed numerically
and approximated by a quadrature formula
\begin{equation}
\label{lr:quadexp}
\rho(r) \approx C \sum_{k=0}^K 
w_k e^{t_k} f_{\alpha \beta}(e^{t_k}) e^{-r^2 e^{t_k}}.
\end{equation}
Here $w_k$ and $e^{t_k}$ are quadrature weights and nodes,
respectively. 
The procedure to compute weights and nodes was
taken from the paper \cite{beylkin-expsumrev-2010}. 
For the reader's convenience we give the formula 
and its derivation in Appendix~\ref{App:rhoIntegral}.

Comparision with equation~\eqref{gaussianExpand} gives
\begin{equation}
\lambda_k = C \, w_k e^{t_k} f_{\alpha \beta}(e^{t_k}),
\qquad \eta_k = e^{t_k}.
\end{equation}

It appears that not many quadrature points (at fixed $r$) are
required to achieve the accuracy of the expansion of order
$10^{-7}$ in the Chebyshev norm due to the hyper-exponential decay. 
Practically, such an accuracy is quite enough for
the physical meaning of the model.
However, $r$ is going to be very small, as 
soon as the finer grids (i.e. for a large number of nodes) are
required for higher precision.
Estimation of upper bound $K$ in sum~\eqref{lr:quadexp} 
which has the logarithmic dependence
is given in Appendix~\ref{App:SlaterKn}.

\subsection{Fast computation of $T(b)$.}
\label{FastTb2DSect}
In this section we describe a numerical scheme 
based on the cross decomposition of two dimensional integrand.
Dimensionality reduction (from three to two dimensions) is achieved by 
means of the separated representation
of Slater density obtained in the previous Section.

Discretization of one-dimensional integrals in \eqref{lrint:2dex}
by some quadrature formula with nodes
$x_i \in [a_x, b_x]$, 
$i = 1,\ldots, n$, $y_j \in [a_y, b_y]$, 
$j = 1,\ldots,m$
and weights $w^{(x)}_i$, $w^{(y)}_j$, 
leads to the approximation
\begin{equation}
\label{IntegralS1D1D}
\iint F(x,y) dx dy \approx \sum_{\tau=1}^q \sigma_{\tau}
\sum_{i=1}^n w^{(x)}_i u_{\tau}(x_i) 
\sum_{j=1}^m w^{(y)}_j g_{\tau}(y_j).
\end{equation}
To get the decomposition~\eqref{lrint:2ddiscr}
we apply \emph{2d-cross} algorithm~\cite{tee-cross-2000, cross2dgit} 
implemented in~\cite{litsarev-cpc-2014}.

A three-dimensional integral $T(b)$ defined in~\eqref{TbDef}
can be reduced to a two-dimensional integral
by means of the decomposition~\eqref{rhoSeparated}
\begin{equation}
T(b)=\sum_{k=0}^{K} \lambda_k \iiint \! \Delta E(x,z - b) \,
e^{-\eta_k x^2} e^{-\eta_k y^2} e^{-\eta_k z^2}  
 dx dy dz
\end{equation}
and analytical evaluation of one-dimensional  Gaussian integral
\begin{equation}
\int^{\infty}_{-\infty} e^{-\eta\, y^2}dy=\sqrt{\frac{\pi}{\eta}},
\end{equation}
\begin{equation}
\label{Tb2DGauss}
T(b)=\sqrt{\pi}
\sum_{k=0}^{K} \frac{\lambda_k}{\sqrt{\eta_k}} 
\iint \! \Delta E(x,z - b) \, e^{-\eta_k x^2}  e^{-\eta_k z^2} dx dz.
\end{equation}
Suppose that $\Delta E(x,z-b)$ has been decomposed as follows
\begin{equation}
\label{DeltaEDecomp}
\Delta E(x, z - b) \approx 
\sum_{\tau=1}^q \sigma_{\tau} u_{\tau}(x) g_{\tau}(z-b).
\end{equation}
Then the integration \eqref{Tb2DGauss} can be reduced to a sequence of
one-dimensional integrations. 
\begin{equation}
\label{Tb1d1dsum}
T(b)=\sqrt{\pi}
\sum_{k=0}^{K} \frac{\lambda_k}{\sqrt{\eta_k}}
\sum_{\tau=1}^{q} \sigma_{\tau} 
I_{\tau k} J_{\tau k}(b),
\end{equation}
\begin{equation}
\label{If1D}
I_{\tau k}=\int_{a_x}^{b_x} \! u_{\tau}(x) e^{-\eta_k x^2} dx,
\end{equation}
\begin{equation}
\label{Ig1D}
J_{\tau k}(b)=
\int_{a_y}^{b_y} \! g_{\tau}(z-b) e^{-\eta_k z^2} dz.
\end{equation}
For the numerical approximation of integrals \eqref{If1D} and
\eqref{Ig1D} 
we use quadrature formula with uniform quadrature nodes (although
any suitable quadrature can be used)
\begin{equation}
\label{NumIka}
I_{\tau k}=
\sum_{i\, \in \,\Omega^x_{\epsilon}(k)}
 w^{(x)}_{i} u_{\tau}(x_i) e^{-\eta_k x_i^2},
 \qquad
 \Omega^x_{\epsilon}(k)=\{\,i  \mid  e^{-\eta_k x_i^2} > \epsilon \, \}
\end{equation}
\begin{equation}
\label{xiPoints}
x_i=-a_x + i \, h_x, \qquad
0 \le i \le 2N_x, \qquad
h_x = a_x/N_x,
\end{equation}
\begin{equation}
\label{Jzmb}
J_{\tau k}(b)=
\sum_{j} 
w^{(z)}_j \,g_{\tau}(z_j-b) e^{-\eta_k z_j^2},
\end{equation}
\begin{equation}
\label{zjPoints}
z_j=-a_z + j \, h_z, \qquad
0 \le j \le 2N_z, \qquad
h_z = a_z/N_z.
\end{equation}
We sample the impact parameter $b$ (which can take only positive
values) with the \emph{same step} $h_z$
\begin{equation}
\label{bPoints}
b_l= l \, h_z, \qquad
0 \le l \le N_z.
\end{equation}
This allows us to introduce a new variable $\tilde z = z - b$
discretized as follows
\begin{equation}
\label{ztildakPoints}
\tilde z_k=-2a_z + k \, h_z, \qquad
0 \le k \le 3N_z,
\end{equation}
and such that for the boundary conditions \eqref{zjPoints},
\eqref{bPoints}, \eqref{ztildakPoints}
\begin{equation}
z_j - b_l = \tilde z_{j-l+N_z}.
\end{equation}
Thus, the approximation problem \eqref{DeltaEDecomp} reduces 
to the low-rank approximation
of the extended $(2N_x + 1) \times (3N_z + 1)$ matrix
\begin{equation}
\label{Eextand}
\Delta E(x_i, \tilde z_j) = 
\sum_{\tau=1}^r \sigma_{\tau} u_{\tau}(x_i) g_{\tau}(\tilde z_j).
\end{equation}
This should be done only once (using the cross approximation
algorithm), and the final approximation of integral \eqref{Jzmb} reads
\begin{equation}
\label{Jakblfast}
J_{\tau k}(b_l) =
\sum_{j \, \in \,\Omega^{\tilde z}_{\epsilon}(k)} 
w^{(\tilde z)}_j \,g_{\tau}(\tilde z_{j-l+N_z}) e^{-\eta_k \tilde z_j^2}.
\end{equation}
The calculation of $T(b)$ can be summarized in 
the following algorithm.
\begin{algorithmic}[1]
\label{alg01}
\FOR{every $\gamma$-shell of the projectile ion}
\STATE compute the decomposition~\eqref{gaussianExpand} for $\rho(r)$
\STATE compute the cross approximation for the matrix $\Delta E(x_i,\tilde z_j)$
defined in~\eqref{Eextand} 
\FOR {$k=0 \ldots K$} 
\FOR {$\tau=1 \ldots q$}
\STATE compute the integral $I_{\tau k}$ defined in~\eqref{NumIka}
\ENDFOR
\ENDFOR
\FOR{every $b_l$ required}
\FOR {$k=0 \ldots K$} 
\FOR {$\tau=1 \ldots q$}
\STATE compute the integral $J_{\tau k}(b_l)$ defined in~\eqref{Jakblfast}
\ENDFOR
\ENDFOR
\STATE compute $T_{\gamma}(b_l)$, equation~\eqref{Tb1d1dsum}
\ENDFOR
\ENDFOR
\end{algorithmic}

Finally, the question is how to compute $T(b)$ for many different values of $b$. The direct summation requires $\mathcal{O}(N^2)$
operations. But a closer look reveals that it is in fact a discrete convolution, 
which can be computed in linear cost. Nevertheless, due to the exponential decay of 
factors~$e^{\eta_k x_i}$ and~$e^{\eta_k \tilde z_j}$
in sums~\eqref{NumIka} and~\eqref{Jakblfast} correspondingly,
there is a very small number of terms, such that
a direct summation after the truncation is much faster. 
This can be easily seen from values of 
parameter~$\theta_x$ defined as
\begin{equation}
\label{thetadef}
\theta_x = \frac{\sum_{k=0}^K \mathcal{N}(\Omega^{x}_{\epsilon}(k)) }{(2N_x +1) (K+1)},
\end{equation}
where $\mathcal{N}(X)$ is the cardinality of set $X$.
For details, please, see Table~\ref{Table2Times} and 
discussion in the following Section~\ref{NumericSect}.

\subsection{Numerical experiments and discussions.}
\label{NumericSect}

Once the analytical expansion~\eqref{rhoSeparated} is obtained,
the full calculation of $T(b)$
consists of two steps (see algorithm in Section~\ref{alg01}): 
calculation of the cross decomposition 
of integrand~\eqref{Tb2DGauss}
on the extended $\tilde z$-mesh~\eqref{ztildakPoints} and
computation of all integrals $I_{\tau k}$ and $J_{\tau k}(b_l)$ for 
all $b_l$ from~\eqref{bPoints}.
There is no need in all the time recalculation of the
cross approximation~\eqref{Eextand}
for the full computation of these integrals.
After it is computed (that is 
$\sigma_{\tau}$, $u_{\tau}(x_i)$ and $g_{\tau}(\tilde z_j)$
are known), the calculation of $I_{\tau k}$ and $J_{\tau k}(b_l)$ starts.

In Table \ref{Table1Ranks} we present 
times $T_{\mbox{\footnotesize{cross}}}$  and ranks $q$
calculated for the energy gains
$\Delta E_{\gamma}(x, \tilde z)$  
corresponding to two ion-atomic collision examples
($Au^{26+}+ O$ and $U^{28+}+ Xe$).
Values of $T_{\mbox{\footnotesize{cross}}}$ are reported for
different sizes of $(x,\tilde z)$-mesh and the relative accuracy
of the cross decomposition.
One can find that $T_{\mbox{\footnotesize{cross}}}$ is 
about $10^{-1}\sim 10^{0}$ second in order of magnitude.
Given that the full computation of integrals $I_{\tau k}$ and $J_{\tau k}(b_l)$ for 
all $b_l$ takes approximately $5000 \cdot 5 \cdot 10^{-3}=25$ seconds
(the average value of column $T_S$ in Table~\ref{Table2Times} 
for $5000$ $b$ values), we can conclude that
the cross approximation becomes then a pre-computing stage 
with a tiny contribution to the total computational time.

An important parameter in sum~\eqref{Eextand} is the rank value $q$. It
determines the complexity of the algorithm (the smaller $q$, the better).
As it follows from the numerical experiments, ranks 
of $\Delta E_{\gamma}(x, \tilde z)$ decomposition are 
small (Table~\ref{Table1Ranks}) against the mesh size.
It means that for real physical systems the cross decomposition
is a very prominent tool. It allows to decrease the complexity of the problem
in practice from $O(n^2)$ elements to $O(q\cdot n)$ elements
where $q \ll n$.

In Table~\ref{Table2Times} we present the program speedup 
for every atomic shell.
In quadrature sum~\eqref{NumIka} all terms less 
then $\epsilon = 10^{-20}$ were thrown out 
for every $x_i$ due to the exponential decay of factors~$e^{-\eta_k x^2_i}$
for every fixed $\eta_k$.
Parameter $\theta_x$ is defined as a relative number of terms 
in total summation of $I_{\tau k}$ over all $k$ and~$i$
(equation~\eqref{NumIka} and~\eqref{thetadef}), 
 above the threshold $\epsilon$. 
As it can be seen from the table, there are only a few percent
of terms to be summed, which 
considerably reduces the sum and speeds up the total calculation.
Parameter~$\theta_x$ decreases
when going from top to bottom of the third column (for one system).
That is why the time~$T_S$ also decreases
while both ranks $K$ and $q$ increase.
The same situation occurs for sum~\eqref{Jakblfast} over~$\tilde z_j$.

Another important question is the full accuracy of the computation.
As it was mentioned above, it consists of two contributions
acquired from the cross approximation and the quadrature summations.
In equations~\eqref{NumIka} and~\eqref{Jakblfast}
values $\sigma_{\tau}$, $u_{\tau}(x_i)$ and $g_{\tau}(\tilde z_j)$
are approximated with the cross accuracy~$\varepsilon_{c}$,
while the quadrature sum is calculated with an error~$\varepsilon_{i}$.
In Table~\ref{Table3IntErr} and Table~\ref{Table4IntErr} we 
provide a numerical example demonstrating 
the actual error for the integral $T(b)$ for small enough 
mesh sizes. We consider the worst case $b=0$, when the integrand
has the most singular behavior (because the Slater density has no 
higher derivatives at $r=0$). 
As it follows from our results the scheme holds the third order
for the Simpson rule, even in the worst singular case. 
In other cases $b>0$ it holds the fourth order.
These results show that the claimed accuracy is achieved
and the numerical scheme holds the order of integration up to the accuracy
of the approximated function.

Finally, we can conclude that usage of the technique based on separated 
representations~\eqref{Tb1d1dsum}
allows to decrease the total time of computing of $T(b)$ 
by a factor of $\sim 10^3$ in comparison to the previous version. In
practice, computational time is 
reduced from several hours to one minute or less on the
same hardware. 

These results were obtained by using of our implementation of the cross 
approximation algorithm 
and the low-rank format of the Slater density~\eqref{lr:quadexp}.
The latest version of the cross decomposition code
implemented in $C$++ can be downloaded from the link~\cite{cross2dgit}.

\section{Conclusions and future work}
\label{ConclSect}
A new and efficient technique for computation of three-dimensional
integrals based on low-rank and separated representations is proposed
for the energy deposition model.
Due to a general form of the integrand, which is a product of 
two functions with cylindrical and spherical symmetries,
this methodology can be applied to many types of integrals
having similar structure.
Such an approach significantly reduces computational time
and allows to achieve a given accuracy (because it uses
the cross approximation and quadrature summations).
The general concept can
be applied to more complicated models 
(like ion-molecular collisions with electron loss and charge-changing 
processes) that lead to multidimensional integrals. 
For the multidimensional case we plan to use fast approximation
techniques based on the tensor train (TT) format~\cite{osel-tt-2011}.

\ack
\label{AcknowledgeSect}
This work was partially supported 
by Russian Science Foundation grant  14-11-00659.

\begin{table}
\caption{
Ranks $q$ of the decomposition~\eqref{Eextand}
calculated by incomplete cross approximation
algorithm~\cite{tee-cross-2000}
for the energy gain $\Delta E(x, \tilde z)$.
Two cases are considered:
collision of $Au^{26+}$ ions with an Oxigen atom
at a collision energy $E=6.5$ MeV/u
and collision of $U^{28+}$ ions with a Xenon atom
at a collision energy $E=2.5$ MeV/u.
The number of $x$-mesh points are taken equal to $2N+1$,
 the number of $\tilde z$-mesh points is taken equal to
$3N+1$ in correspondence to 
 equations~\eqref{xiPoints} and~\eqref{ztildakPoints}, $a_x=a_z=8$.
The relative error of the approximation in the Frobenius norm
$\varepsilon$ and $N$ are placed in the bottom line as 
a couple~$(\varepsilon, N)$. They correspond to different numerical tests.
Calculations were carried out on $1.3$ GHz Intel Core i5 processor.
Column $T_{\mbox{\footnotesize{cross}}}$
corresponds to the time the cross algorithm takes.
The numerical results confirm almost linear scaling of the approach in $N$.
}
\label{Table1Ranks}
\centering
\begin{tabular}{cccccccc}
\hline
\hline
System & $\gamma$-Shell & $q$ & $T_{\mbox{\footnotesize{cross}}}$ (sec)
  & $q$ & $T_{\mbox{\footnotesize{cross}}}$ (sec) & 
  $q$ & $T_{\mbox{\footnotesize{cross}}}$ (sec) \\
\hline
$Au^{26+}+ O$ & $4df^{17}$ & $13$ & $0.21$ 
& $21$ & $0.42$ &  24 & 2.41 \\
 & $4sp^{8}$ & $13$ & $0.21$ & $21$ & $0.33$ & 24 & 2.40 \\
 & $3d^{10}$ & $14$ & $0.19$ & $22$ & $0.42$ & 26 & 2.58 \\
 & $3sp^{8}$ & $16$ & $0.25$ & $24$ & $0.54$ & 29 & 2.64 \\
 & $2sp^{8}$ & $17$ & $0.28$ & $25$ & $0.56$ & 30 & 2.71 \\
 & $1sp^{2}$ & $17$ & $0.27$ & $25$ & $0.56$ & 30 & 2.70 \\
\hline
$U^{28+}+ Xe$ & $5sp^{4}$ & $14$ & $0.20$ &  
 $22$ & $0.50$ & 26 & 2.17 \\
 & $4df^{24}$ & $15$ & $0.23$ & $24$ & $0.52$ & 27 & 2.58 \\
 & $4sp^{8}$ & $17$ & $0.28$ & $25$ & $0.55$ & 30 & 2.69 \\
 & $3d^{10}$ & $17$ & $0.27$ & $25$ & $0.54$ & 30 & 2.77 \\
 & $3sp^{8}$ & $17$ & $0.27$ & $25$ & $0.55$ & 30 & 2.75 \\
 & $2sp^{8}$ & $17$ & $0.26$ & $25$ & $0.54$ & 30 & 2.71 \\
 & $1sp^{2}$ & $17$ & $0.26$ & $25$ & $0.55$ & 30 & 2.69 \\
\hline
 &  & \multicolumn{2}{l}{($10^{-6}$, $1024$)} &
 \multicolumn{2}{l}{($10^{-9}$, $1024$)} & 
 \multicolumn{2}{l}{($10^{-9}$, $4096$)} \\
\hline
\hline
\end{tabular}
\end{table}

\begin{table}
\caption{
Timings to compute  $T(b)$ at fixed $b$ are
presented for two cases: the DEPOSIT code (old) $T_D$
and the code based on separated 
representations~\eqref{Tb1d1dsum} $T_S$.
Collision systems are the same as in Table~\ref{Table1Ranks}.
In the third column $K$ labeles maximum value of index 
in expansion~\eqref{rhoSeparated}.
The mesh~\eqref{QuadSlaterMesh} for radial 
density is taken with $a_t=-3$, $b_t=45$, $h_t=(b_t - a_t)/250$.
The meaning of parameter $\theta_x$ is explained in
Section~\ref{NumericSect}.
Calculations were carried out
for the relative accuracy of the cross 
decomposition $\varepsilon=10^{-7}$ and 
$[-8,8]\otimes[-16,8]$ mesh with
$4097 \times 6145$ points.
The last column shows speedup of the program.
}
\label{Table2Times}
\centering
\begin{tabular}{ccccccc}
\hline
\hline
System & $\gamma$-Shell & $K$& $\theta_x \times 10^{-2}$ & $T_S$ ($\times10^{-3}$ sec) &
 $T_{D}$ (sec) & $T_{D}/T_S$  \\
\hline
$Au^{26+}+ O$ & $4df^{17}$ & $74$ & $9.1$ & $7.94$ & $3.89$ & $490$  \\
 & $4sp^{8}$ & $69$ & $6.0$ & $4.92$ & $3.83$ & $778$  \\
 & $3d^{10}$ & $73$ & $4.3$ & $3.59$ & $3.88$ & $1080$  \\
 & $3sp^{8}$ & $72$ & $3.9$ & $3.81$ & $3.82$ & $1003$  \\ 
 & $2sp^{8}$ & $107$ & $1.5$ & $2.42$ & $3.86$ & $1592$  \\
 & $1sp^{2}$ & $209$ & $0.4$ & $1.24$ & $3.88$ & $3120$  \\  
\hline
$U^{28+}+ Xe$ & $5sp^{4}$ & $62$ & $13.1$ & $10.1$ & $3.94$ & $390$ \\
 & $4df^{24}$ & $70$ & $6.5$ & $6.05$ & $3.90$ & $644$ \\
 & $4sp^{8}$ & $67$ & $5.1$ & $5.00$ & $3.94$ & $788$ \\
 & $3d^{10}$ & $71$ & $3.6$ & $3.88$ & $3.92$ & $1011$ \\
 & $3sp^{8}$ & $70$ & $3.4$ & $3.52$ & $3.90$ & $1106$ \\
 & $2sp^{8}$ & $105$ & $1.3$ & $1.99$ & $3.87$ & $1945$ \\
 & $1sp^{2}$ & $207$ & $0.3$ & $1.04$ & $3.88$ & $3723$ \\
\hline
\hline
\end{tabular}
\end{table}

\begin{table}
\caption{
Convergence of integrals $T(b)$ for $4df^{17}$, $4sp^{8}$ and $3d^{10}$ shells
of $Au^{26+}+ O$ at $6.5$ MeV/u
on two-dimensional mesh
$[-10,10]\otimes[-20,10]$ with $(2N+1) \times (3N+1)$ points 
for different~$N$ (first column), see 
equations~\eqref{xiPoints} and~\eqref{ztildakPoints}.
Simpson weights $w^{x}_{i}$, $w^{(\tilde z)}_j$ are used in~\eqref{NumIka} and~\eqref{Jakblfast}.
For radial density the mesh~\eqref{QuadSlaterMesh}
is taken with parameters $a_t=-3$, $b_t=45$, $h_t=(b_t - a_t)/650$.
Calculations were carried out
for the relative accuracy of the cross 
decomposition $\varepsilon_{c}=10^{-12}$ and 
fixed value of parameter $b_0=0.0$.
Last column shows the rellative error $\varepsilon_i$.
Extrapolated value of integral is calculated by Romberg method (with Richardson 
extrapolation)~\cite{stoer2002},~p.~161.
The order of scheme~$p_{e}$ is defined by Aitken rule~\cite{stoer2002},~p.~344.
}
\label{Table3IntErr}
\centering
\begin{tabular}{crcccc}
\hline
\hline
$\gamma$-Shell & $N$ & $q$ & $T(b_0)$ & $p_{e}$ & $\varepsilon_i$ \\
\hline
$4df^{17}$
&     $2048$ & $30$ & $177.131769802401$ &               & $2.1 \cdot 10^{-7}$ \\
&     $4096$ & $33$ & $177.131804304375$ &               & $1.2 \cdot 10^{-8}$ \\
&     $8192$ & $35$ & $177.131806364504$ & $4.07$ & $7.7 \cdot 10^{-10}$ \\
&   $16384$ & $37$ & $177.131806491972$ & $4.01$ & $4.8 \cdot 10^{-11}$ \\
&   $32768$ & $39$ & $177.131806499918$ & $4.00$ & $3.0 \cdot 10^{-12}$ \\
&   $65536$ & $40$ & $177.131806500407$ & $4.02$ & $2.3 \cdot 10^{-13}$ \\
& \multicolumn{2}{r}{\textit{{\small extrapolated}}}  & $177.131806500448$ & & \\
\hline
$4sp^{8}$ 
&     $2048$ & $31$ & $165.507815905465$ &             & $2.2 \cdot 10^{-7}$  \\
&     $4096$ & $33$ & $165.507850781567$ &             & $1.3 \cdot 10^{-8}$ \\
&     $8192$ & $35$ & $165.507852865842$ & $4.06$ & $8.3 \cdot 10^{-10}$ \\
&   $16384$ & $37$ & $165.507852994821$ & $4.01$ & $5.2 \cdot 10^{-11}$ \\
&   $32768$ & $39$ & $165.507853002863$ & $4.00$ & $3.2 \cdot 10^{-12}$ \\
&   $65536$ & $40$ & $165.507853003364$ & $4.01$ & $2.1 \cdot 10^{-13}$ \\
& \multicolumn{2}{r}{\textit{{\small extrapolated}}}  & $165.507853003399$ &  & \\
\hline
$3d^{10}$ 
&     $2048$ & $33$ & $407.589200892012$ &             & $5.0 \cdot 10^{-7}$  \\
&     $4096$ & $35$ & $407.589382823491$ &             & $5.2 \cdot 10^{-8}$  \\
&     $8192$ & $38$ & $407.589401536350$ & $3.28$ & $5.8 \cdot 10^{-9}$  \\
&   $16384$ & $40$ & $407.589403612947$ & $3.17$ & $6.7 \cdot 10^{-10}$  \\
&   $32768$ & $42$ & $407.589403853285$ & $3.11$ & $8.0 \cdot 10^{-11}$  \\
&   $65536$ & $44$ & $407.589403882005$ & $3.06$ & $9.8 \cdot 10^{-12}$  \\
& $131072$ & $46$ & $407.589403885498$ & $3.04$ & $1.2 \cdot 10^{-12}$  \\
& $262144$ & $47$ & $407.589403885925$ & $3.03$ & $1.9 \cdot 10^{-13}$  \\
& \multicolumn{2}{r}{\textit{{\small extrapolated}}}  & $407.589403886002$ &  & \\
\hline
\hline
\end{tabular}
\end{table}

\begin{table}
\caption{
Convergence of integrals $T(b)$ for
$3sp^{8}$, $2sp^{8}$ and $1s^{2}$ shells
of $Au^{26+}+ O$ at $6.5$ MeV/u
on two-dimensional mesh
$[-10,10]\otimes[-20,10]$ with $(2N+1) \times (3N+1)$ points 
for different~$N$ (first column), see 
equations~\eqref{xiPoints} and~\eqref{ztildakPoints}.
Simpson weights $w^{x}_{i}$, $w^{(\tilde z)}_j$ are used in~\eqref{NumIka} and~\eqref{Jakblfast}.
For radial density the mesh~\eqref{QuadSlaterMesh}
is taken with parameters $a_t=-3$, $b_t=45$, $h_t=(b_t - a_t)/650$.
Calculations were carried out
for the relative accuracy of the cross 
decomposition $\varepsilon_{c}=10^{-12}$ and 
fixed value of parameter $b_0=0.0$.
Last column shows the rellative error $\varepsilon_i$.
Extrapolated value of integral is calculated by Romberg method (with Richardson 
extrapolation)~\cite{stoer2002},~p.~161.
The order of scheme~$p_{e}$ is defined by Aitken rule~\cite{stoer2002},~p.~344.
}
\label{Table4IntErr}
\centering
\begin{tabular}{crcccc}
\hline
\hline
$\gamma$-Shell & $N$ & $q$ & $T(b_0)$ & $p_{e}$ & $\varepsilon_i$ \\
\hline
$3sp^{8}$ 
&       $2048$ & $35$ & $53.0906497215656$ &               & $3.0 \cdot 10^{-5}$  \\
&       $4096$ & $38$ & $53.0920135580097$ &               & $4.6 \cdot 10^{-6}$  \\
&       $8192$ & $41$ & $53.0922247714446$ & $2.69$ & $6.1 \cdot 10^{-7}$  \\
&     $16384$ & $44$ & $53.0922529097596$ & $2.91$ & $7.7 \cdot 10^{-8}$  \\
&     $32768$ & $47$ & $53.0922564862273$ & $2.98$ & $9.7 \cdot 10^{-9}$  \\
&     $65536$ & $50$ & $53.0922569351057$ & $2.99$ & $1.2 \cdot 10^{-9}$  \\
&   $131072$ & $52$ & $53.0922569912665$ & $3.00$ & $1.5 \cdot 10^{-10}$  \\
&   $262144$ & $54$ & $53.0922569982873$ & $3.00$ & $1.9 \cdot 10^{-11}$  \\
&   $524288$ & $56$ & $53.0922569991611$ & $3.01$ & $2.4 \cdot 10^{-12}$  \\
& $1048576$ & $58$ & $53.0922569992692$ & $3.01$ & $4.0 \cdot 10^{-13}$  \\
& \multicolumn{2}{r}{\textit{{\small extrapolated}} }  & $53.0922569992903$ &  & \\
\hline
$2sp^{8}$ 
&       $2048$ & $36$ & $4.97891259072213$ &  & $2.3 \cdot 10^{-4}$  \\
&       $4096$ & $39$ & $4.97986325621640$ &  & $3.5 \cdot 10^{-5}$  \\
&       $8192$ & $42$ & $4.98001545713255$ & $2.64$ & $4.7 \cdot 10^{-6}$  \\
&     $16384$ & $45$ & $4.98003568931593$ & $2.91$ & $5.9 \cdot 10^{-7}$  \\
&     $32768$ & $48$ & $4.98003825660591$ & $2.98$ & $7.4 \cdot 10^{-8}$  \\
&     $65536$ & $51$ & $4.98003857868963$ & $2.99$ & $9.2 \cdot 10^{-9}$  \\
&   $131072$ & $53$ & $4.98003861898554$ & $3.00$ & $1.2 \cdot 10^{-9}$  \\
&   $262144$ & $56$ & $4.98003862402352$ & $3.00$ & $1.4 \cdot 10^{-10}$  \\
&   $524288$ & $58$ & $4.98003862465325$ & $3.00$ & $1.8 \cdot 10^{-11}$  \\
& $1048576$ & $60$ & $4.98003862473183$ & $3.00$ & $2.3 \cdot 10^{-12}$  \\
& $2097152$ & $61$ & $4.98003862474152$ & $3.02$ & $3.5 \cdot 10^{-13}$  \\
& \multicolumn{2}{r}{\textit{{\small extrapolated}}}  & $4.98003862474327$ &  & \\
\hline
$1sp^{2}$ 
&       $2048$ & $36$ & $0.122305706797573$ &  & $8.3 \cdot 10^{-3}$  \\
&       $4096$ & $39$ & $0.123229108221677$ &  & $8.3 \cdot 10^{-4}$  \\
&       $8192$ & $42$ & $0.123322068141309$ & $3.31$ & $7.9 \cdot 10^{-5}$  \\
&     $16384$ & $45$ & $0.123330850494328$ & $3.40$ & $8.2 \cdot 10^{-6}$  \\
&     $32768$ & $48$ & $0.123331746174083$ & $3.29$ & $9.1 \cdot 10^{-7}$  \\
&     $65536$ & $51$ & $0.123331845477160$ & $3.17$ & $1.1 \cdot 10^{-7}$  \\
&   $131072$ & $53$ & $0.123331857117543$ & $3.09$ & $1.3 \cdot 10^{-8}$  \\
&   $262144$ & $56$ & $0.123331858525395$ & $3.05$ & $1.6 \cdot 10^{-9}$  \\
&   $524288$ & $58$ & $0.123331858698474$ & $3.02$ & $2.0 \cdot 10^{-10}$  \\
& $1048576$ & $59$ & $0.123331858719929$ & $3.01$ & $2.5 \cdot 10^{-11}$  \\
& $2097152$ & $61$ & $0.123331858722600$ & $3.01$ & $3.1 \cdot 10^{-12}$  \\
& \multicolumn{2}{r}{\textit{{\small extrapolated}}}  & $0.123331858722980$ &  & \\
\hline
\hline
\end{tabular}
\end{table}

\appendix
\section{Physical model}
\label{App:PhysModel}
In this section we introduce functions and parameters involved into 
the definition of deposited energy integral~\eqref{TbDef}
\begin{equation}
\label{App:Tb}
T_{\gamma}(b)=\iiint \! \Delta E_{\gamma}(p) 
\rho_{\gamma}(x,y,z) dx dy dz,
\end{equation}
\begin{equation}
p=\sqrt{(z - b)^2 + x^2},
\end{equation}
where the integration is done over the whole coordinate space.
Integral~\eqref{App:Tb} is written for an atomic shell 
with \textit{principal} quantum number $n$ and \textit{orbital}
quantum number $l$ labeled by $\gamma=nl$.

The energy gain $\Delta E_{\gamma}(p)$ is a kinetic energy
deposited to the projectile's $\gamma$-shell by the field~$U(R)$
of the target atom.
\begin{equation}
\label{App:dEgain}
\Delta E_{\gamma}(p)=
\Delta E^{<}_{\gamma}(p)\, n_f(v_{\gamma} - \vartheta) + 
\Delta E^{>}_{\gamma}(p)\, n_f(\vartheta - v_\gamma),
\end{equation}
\begin{equation}
n_f(x)=\frac{1}{e^{-kx} +1},
\end{equation}
where $k=3$ by default and is an input parameter of the model.
Expression~\eqref{App:dEgain} consists of two terms corresponding to low 
\begin{equation}
\label{ELow}
\Delta E^{<}_{\gamma}(p)=
\frac{ {\xi}_{\gamma} }{ p+ \zeta_{\gamma} }
\sum^{3}_{i=1} \phi_i F(\chi_i p)
\end{equation}
and high 
\begin{equation}
\label{EHigh}
\Delta E^{>}_{\gamma}(p)=
\frac{ {\nu}_{\gamma} }{ p^2 + \mu_{\gamma} }
\left(\sum^{3}_{i=1} \phi_i F(\chi_i p)\right)^2 \!\!,
\end{equation}
energies. Function
\begin{equation}
F(x)=x K_1(x)=\int_{0}^{\infty} \!\! e^{-\sqrt{x^2+y^2}}dy
\end{equation}
is defined in terms of the
\textit{modified Bessel function of the second kind}
$K_1(x)$ (see~\cite{abramowitz-stegun}, p.~375, Eq.~9.6.2) with asympthotical behavior
(\cite{abramowitz-stegun}, p.~378, Eq.~9.8.3 and p.~379, Eq.~9.8.7)
\begin{equation}
F(x\to 0_{+})=1+\left(\frac{1}{2} \ln \frac{x}{2} + 0.03860786\right)x^2 + \mathcal{O}(x^3).
\end{equation}

Fitting parameters $\phi_i$, $\chi_i$ of the atomic field $U(R)$ are obtained from the 
Dirac-Hartree-Fock-Slater calculations~\cite{PhysRevA.36.467}. The distance
between the center of the field and the projectile electron of 
$\gamma$-shell is labeled by $R$.
Atomic field is given in the Yukawa potential form
\begin{equation}
U(R)=-\frac{Z}{R} \sum_{i=1}^{3} \phi_i e^{-\chi_i R},
\end{equation}
where $Z$ is the nuclear charge of the target atom.

Parameter $\vartheta$ is a relative velocity of the projectile.
It is related with the collision energy of the ion-atomic system.
The rest parameters $v_\gamma$, $\xi_\gamma$, 
$\zeta_\gamma$, $\nu_\gamma$ and $\mu_\gamma$ 
concern to fundamental properties of the projectile ion and the target atom
(such as binding energy, average velocity, mean radius of the shell,
atomic radius and charge).
They should be considered as positive constants in the integral~\eqref{App:Tb} 
for a given ion-atomic system for all $b$.
Detailed description of how to calculate them 
can be found in~\cite{litsarev-cpc-2013}.
Examples of input files with the original code
can be downloaded from link~\cite{depositgit}.

\section{Inverse Laplace transform sources} 
\label{App:Laplace}
For integer $\alpha$ and real positive $\beta$ inverse Laplace
transform $f_{\alpha \beta}(t)$ of $F_{\alpha \beta}(s)$
from equation~\eqref{FabsqS} may be calculated analytically
and expressed via \emph{the Kummer's confluent hypergeometric function}
$M(a,b;z)$ (\cite{abramowitz-stegun}, chapter~13) as follows
\begin{equation}
f_{\alpha\beta}(t)=
\frac{M\left(1+\frac{\alpha}{2},\frac{1}{2},-\frac{\beta^2}{t}\right)}
{t^{1+\frac{\alpha}{2}} \Gamma\left(-\frac{\alpha}{2} \right)} 
-2\,\beta\,\frac{M\left(\frac{3+\alpha}{2},\frac{3}{2},-\frac{\beta^2}{t}\right)}
{t^{\frac{3+\alpha}{2}} \Gamma\left(-\frac{1+\alpha}{2} \right)},
\end{equation}
where
\begin{equation}
M(a,b;z)=1+\frac{a}{b}\frac{z}{1!}+
\frac{a(a+1)}{b(b+1)}\frac{z^2}{2!}+\ldots
\end{equation}
and $\Gamma(x)$ is \emph{the Gamma function}.

Below we present $f_{\alpha \beta}(t)$ explicitly
for most practically usefull cases ($\alpha=0,1,\ldots, 6$).
Due to the difference of normalization conditions
in spherical and Cartesian coordinates
for the Slater density~\eqref{rhoDef}
\begin{equation}
\label{RhoSpher}
\rho(r)= N_{\gamma}\frac{(2\beta)^{2\mu+1}}{\Gamma({2\mu+1})} r^{2 \mu} e^{-2\beta r},
\end{equation}
parameter $\alpha$ in~\eqref{rhoDef} is related to parameter $\mu$
in~\eqref{RhoSpher} as follows
\begin{equation}
\alpha = 2 \mu - 2.
\end{equation}
The number of electrons in the shell $\gamma$ is labeled as~$N_{\gamma}$.
Parameter $\mu$ is greater or equal to unity.  It is an integer or half-integer depending on
\emph{the principal quantum number} $n$ and
\emph{the orbital quantum number} $l$ of the atomic shell.
Details can be found in~\cite{slater1960,shevelko1993}.
For example, $\mu_{1s^2}=1$, $\alpha=0$;
$\mu_{2sp^8}=2$, $\alpha = 2$;
$\mu_{4d^{10}}=3.5$, $\alpha = 5$.
Finally,
\begin{equation}
f_{0 \beta}(t)= \frac{g_0\left(t/\beta^2 \right)}{ \sqrt{\pi} \, \beta^2},
\quad g_0(t)= \frac{e^{-\frac{1}{t}}}{t^{3/2}}
\end{equation}
\begin{equation}
f_{1 \beta}(t)= \frac{g_1\left(t/\beta^2\right)}{ 2\!\sqrt{\pi} \,\beta^3},
\quad
g_1(t) = -\frac{e^{-\frac{1}{t}}}{t^{3/2}}\left(1-  \frac{2}{t} \right)
\end{equation}
\begin{equation}
f_{2 \beta}(t)=\frac{3\, g_2\left(t/ \beta^2\right)} {2\!\sqrt{\pi} \, \beta^4},
\quad
g_2(t)=-\frac{e^{-\frac{1}{t}}} {t^{5/2}} \left(1-  \frac{2}{3t} \right)
\end{equation}
\begin{equation}
f_{3 \beta}(t)= \frac{3 \, g_3\left(t/\beta^2\right)}{4\!\sqrt{\pi}\,\beta^5},
\quad
g_3(t)=\frac{e^{-\frac{1}{t}}}{t^{5/2}}
\left(1 -  \frac{4}{t} + \frac{4}{3t^2} \right)
\end{equation}
\begin{equation}
f_{4 \beta}(t)=\frac{15 \,g_4\left(t/\beta^2\right)}{4 \! \sqrt{\pi} \, \beta^6},
\quad
g_{4}(t)=\frac{e^{-\frac{1}{t}}}{t^{7/2}}
\left(1 -  \frac{4}{3t} + \frac{4}{15t^2} \right)
\end{equation}
\begin{equation}
f_{5 \beta}(t)=\frac{15\, g_5\left(t/\beta^2 \right)}{8 \! \sqrt{\pi} \, \beta^{7}},
\quad
g_5(t)=-\frac{e^{-\frac{1}{t}}}{t^{7/2}} \left(1-  \frac{6}{t} +\frac{4}{t^2} - \frac{8}{15t^3} \right)
\end{equation}
\begin{equation}
f_{6 \beta}(t)=\frac{105\, g_6\left(t/\beta^2\right)}{8\! \sqrt{\pi}\, \beta^8},
\quad
g_6(t) = -\frac{e^{-\frac{1}{t}}}{ t^{9/2}}
\left(1- \frac{2}{t} +\frac{4}{5t^2}- \frac{8}{105t^3}  \right)
\end{equation}
For practical reasons higher values of $\alpha$ are not necessarily
due to the limitation of shell filling with electrons.

\section{Quadrature formula for the Laplace integral} 
\label{App:rhoIntegral}
To obtain the decomposition~\eqref{gaussianExpand}
for given $\alpha$ and $\beta$
we make a substitution $s \rightarrow s^2$ into the equation~\eqref{FabsqS}
\begin{equation}
F_{\alpha \beta}(s^2)=s^{\alpha} e^{-2\beta s}=
\int^{\infty}_{0} e^{-s^2 x}   f_{\alpha \beta}(x) \,dx,
\end{equation}
then introduce another variable  $x=e^{t}$ in the integral
\begin{equation}
\label{IntegralExpt}
F_{\alpha \beta}(s^2)=s^{\alpha} e^{-2\beta s}=
\int^{\infty}_{-\infty} e^{-s^2 e^{t}+t}   f_{\alpha \beta}(e^{t}) dt.
\end{equation}
The function under the integral \eqref{IntegralExpt}
has exponential decay both in the spatial and frequency domains.
Therefore the truncated trapezoidal  rule 
gives the optimal convergence
rate. It turns out to the final approximation of the form
\begin{equation}
\label{IntegralGaussWeights}
 F_{\alpha \beta}(s^2) \approx \sum_{k=0}^K 
w_k e^{t_k} f_{\alpha \beta}(e^{t_k}) e^{-s^2 e^{t_k}}
\end{equation}
with trapezoidal weights $w_k$ and the integrand values in
the nodes $e^{-s^2 e^{t_k} + t_k} f_{\alpha \beta}(e^{t_k})$.
The Gaussian-like part is split out in a separate factor
in correspondence with decomposition~\eqref{gaussianExpand}.
Parameters of the formula
\begin{equation}
\label{QuadSlaterMesh}
t_k = a_t + k h_t, \quad
h_t = (b_t - a_t)/K
\end{equation}
have to be selected in such a way that the resulting quadrature
formula approximates the integral for a wide range of parameter~$s$. 
Typically, the choice $a_t \gtrsim -3$, $b_t \lesssim 45$, and $K\sim 250$
gives good accuracy ($\le 10^{-7}$).
As an example, in Table~\ref{Table2Times} the required number of terms
in sum~\eqref{IntegralGaussWeights} is presented.
Accurate error analysis can be found in \cite{beylkin-expsumrev-2010}.

\section{Convergence rate of $K(N)$ for the Slater density series}
\label{App:SlaterKn}

In this section we estimate an upper bound $K$ in sum~\eqref{lr:quadexp} 
which has the logarithmic dependence on the mesh size~$N$.
We follow the proof given in paper~\cite{Khoromskij2006Kn}
for Lemma~4.3. In this Lemma a Slater-type 
function $\rho(\sqrt{s})$ from~eq.~\eqref{FabsqS} is considered for $\alpha = 0$.
Below, this function is considered for integer $\alpha$.

The integrand in~\eqref{FabsqS}
after the change of variables $t=e^{x}$
\begin{equation}
P_{\alpha \beta}(x)= e^{- s e^{x} + x} f_{\alpha \beta} (e^{x})
\end{equation}
has the following decay on the real axis 
(skipping the numerical factor before the exponent)
\begin{equation}
P_{\alpha \beta}(x) \approx e^{- s e^{x} -c_1 x}, 
\quad \mbox{as~} x \to \infty, \quad
c_1=\frac{\alpha + 1 - (\alpha \bmod 2)}{2},
\end{equation}
\begin{equation}
P_{\alpha \beta}(x) \approx e^{-\beta^2 e^{|x|} + c_2 |x|}, 
\quad \mbox{as~}x \to -\infty, \quad
c_2=\alpha +\frac{1}{2}.
\end{equation}
This immediately implies expression~(5.3) from~\cite{Khoromskij2006Kn}
for $b=\min(\beta^2, s)$, $b$ is taken in the notation of~\cite{Khoromskij2006Kn}. 
Following then the statement~I of Lemma~4.3 we may conclude,
that $K=\mathcal O(|\log \varepsilon| (|\log \varepsilon| + \log1/b))$
with the error~$\varepsilon$ of the approximation and $b\sim1/N^2$.

\bibliographystyle{wileyj}
\bibliography{refs,misha} 

\begin{thebibliography}{10}
\providecommand{\url}[1]{\texttt{#1}}
\providecommand{\urlprefix}{URL }
\expandafter\ifx\csname urlstyle\endcsname\relax
  \providecommand{\doi}[1]{doi:\discretionary{}{}{}#1}\else
  \providecommand{\doi}{doi:\discretionary{}{}{}\begingroup
  \urlstyle{rm}\Url}\fi

\bibitem{beylkin-2002}
Beylkin G, Mohlenkamp MJ. Numerical operator calculus in higher dimensions.
  \emph{Proc. Nat. Acad. Sci. USA}  2002; \textbf{99}(16):10\,246--10\,251,
  \doi{10.1073/pnas.112329799}.

\bibitem{vkhs-2el-2013}
Khoromskaia V, Khoromskij BN, Schneider R. Tensor-structured factorized
  calculation of two-electron integrals in a general basis. \emph{SIAM J. Sci.
  Comput.}  2013; \textbf{35}(2):A987--A1010, \doi{10.1137/120884067}.

\bibitem{kolda-review-2009}
Kolda TG, Bader BW. Tensor decompositions and applications. \emph{SIAM Review}
  2009; \textbf{51}(3):455--500, \doi{10.1137/07070111X}.

\bibitem{tee-cross-2000}
Tyrtyshnikov EE. Incomplete cross approximation in the mosaic--skeleton method.
  \emph{Computing}  2000; \textbf{64}(4):367--380, \doi{10.1007/s006070070031}.

\bibitem{bebe-2000}
Bebendorf M. Approximation of boundary element matrices. \emph{Numer. Mathem.}
  2000; \textbf{86}(4):565--589, \doi{10.1007/pl00005410}.

\bibitem{ost-tucker-2008}
Oseledets IV, Savostianov DV, Tyrtyshnikov EE. Tucker dimensionality reduction
  of three-dimensional arrays in linear time. \emph{SIAM J. Matrix Anal. Appl.}
   2008; \textbf{30}(3):939--956, \doi{10.1137/060655894}.

\bibitem{crossconv2015}
Rakhuba MV, Oseledets IV. Fast multidimensional convolution in low-rank formats
  via cross approximation. \emph{SIAM J. Sci. Comput.}  2015; :in press.

\bibitem{ot-tt-2009}
Oseledets IV, Tyrtyshnikov EE. Breaking the curse of dimensionality, or how to
  use {SVD} in many dimensions. \emph{SIAM J. Sci. Comput.}  2009;
  \textbf{31}(5):3744--3759, \doi{10.1137/090748330}.

\bibitem{osel-tt-2011}
Oseledets IV. Tensor-train decomposition. \emph{SIAM J. Sci. Comput.}  2011;
  \textbf{33}(5):2295--2317, \doi{10.1137/090752286}.

\bibitem{parafac1970}
Harshman RA. Foundation of the parafac procedure: Model and conditions for an
  explanatory multimode factor analysis. \emph{UCLA Working papers in
  phonetics}  1970; \textbf{16}:1--84.

\bibitem{osel-constr-2013}
Oseledets IV. Constructive representation of functions in low-rank tensor
  formats. \emph{Constr. Appr.}  2013; \textbf{37}(1):1--18,
  \doi{10.1007/s00365-012-9175-x}.
  \urlprefix\url{http://pub.inm.ras.ru/pub/inmras2010-04.pdf}.

\bibitem{beylkin-high-2005}
Beylkin G, Mohlenkamp MJ. {Algorithms for numerical analysis in high
  dimensions}. \emph{SIAM J. Sci. Comput.}  2005; \textbf{26}(6):2133--2159.

\bibitem{khor-low-rank-kron-P1-2006}
Hackbusch W, Khoromskij BN. Low-rank {Kronecker-product} approximation to
  multi-dimensional nonlocal operators. {I}. {Separable} approximation of
  multi-variate functions. \emph{Computing}  2006; \textbf{76}(3-4):177--202,
  \doi{10.1007/s00607-005-0144-0}.

\bibitem{khor-low-rank-kron-P2-2006}
Hackbusch W, Khoromskij BN. Low-rank {Kronecker-product} approximation to
  multi-dimensional nonlocal operators. {II}. {HKT} representation of certain
  operators. \emph{Computing}  2006; \textbf{76}(3-4):203--225,
  \doi{10.1007/s00607-005-0145-z}.

\bibitem{khor-acc-2010}
Khoromskij BN. Fast and accurate tensor approximation of multivariate
  convolution with linear scaling in dimension. \emph{J. Comp. Appl. Math.}
  2010; \textbf{234}(11):3122--3139, \doi{10.1016/j.cam.2010.02.004}.

\bibitem{khor-qtt-2011}
Khoromskij BN. $\mathcal{O}(d \log n)$--{Quantics} approximation of {$N$--$d$}
  tensors in high-dimensional numerical modeling. \emph{Constr. Appr.}  2011;
  \textbf{34}(2):257--280, \doi{10.1007/s00365-011-9131-1}.

\bibitem{khor-prec-2009}
Khoromskij BN. Tensor-structured preconditioners and approximate inverse of
  elliptic operators in {$\mathbb{R}^d$}. \emph{Constr. Approx}  2009;
  (30):599--620, \doi{10.1007/s00365-009-9068-9}.

\bibitem{khor-ml-2009}
Khoromskij BN, Khoromskaia V. Multigrid accelerated tensor approximation of
  function related multidimensional arrays. \emph{SIAM J. Sci. Comput.}  2009;
  \textbf{31}(4):3002--3026, \doi{10.1137/080730408}.

\bibitem{mpi-chem3d-2009}
Khoromskij BN, Khoromskaia V, Chinnamsetty SR, Flad HJ. Tensor decomposition in
  electronic structure calculations on {3D Cartesian grids}. \emph{J. Comput.
  Phys.}  2009; \textbf{228}(16):5749--5762, \doi{10.1016/j.jcp.2009.04.043}.

\bibitem{bohr1915}
Bohr N. On the decrease of velocity of swiftly moving electrified particles in
  passing through matter. \emph{Philosophical Magazine Series 6}  1915;
  \textbf{30}(178):581--612, \doi{10.1080/14786441008635432}.

\bibitem{RussekMeli1970}
Russek A, Meli J. Ionization phenomena in high-energy atomic collisions.
  \emph{Physica}  1970; \textbf{46}(2):222--243.

\bibitem{cocke_pra1979}
Cocke CL. Production of highly charged low-velocity recoil ions by heavy-ion
  bombardment of rare-gas targets. \emph{Phys. Rev. A}  Sep 1979;
  \textbf{20}:749--758, \doi{10.1103/PhysRevA.20.749}.

\bibitem{litsarev-jpb-2008}
Shevelko VP, Litsarev MS, Tawara H. Multiple ionization of fast heavy ions by
  neutral atoms in the energy deposition model. \emph{Journal of Physics B:
  Atomic, Molecular and Optical Physics}  2008; \textbf{41}(11):115\,204.

\bibitem{litsarev-jpb-2010}
Shevelko VP, Kato D, Litsarev MS, Tawara H. The energy-deposition model:
  electron loss of heavy ions in collisions with neutral atoms at low and
  intermediate energies. \emph{Journal of Physics B: Atomic, Molecular and
  Optical Physics}  2010; \textbf{43}(21):215\,202.

\bibitem{litsarev-nimb-2009}
Song MY, Litsarev MS, Shevelko VP, Tawara H, Yoon JS. Single- and
  multiple-electron loss cross-sections for fast heavy ions colliding with
  neutrals: Semi-classical calculations. \emph{Nuclear Instruments and Methods
  in Physics Research Section B: Beam Interactions with Materials and Atoms}
  2009; \textbf{267}(14):2369 -- 2375.

\bibitem{litsarev-jpb-2009}
Shevelko VP, Litsarev MS, Song MY, Tawara H, Yoon JS. Electron loss of fast
  many-electron ions colliding with neutral atoms: possible scaling rules for
  the total cross sections. \emph{Journal of Physics B: Atomic, Molecular and
  Optical Physics}  2009; \textbf{42}(6):065\,202.

\bibitem{litsarev-springer-2012}
Shevelko V, Litsarev M, Stohlker T, Tawara H, Tolstikhina I, Weber G. Electron
  loss and capture processes in collisions of heavy many-electron ions with
  neutral atoms. \emph{Atomic Processes in Basic and Applied Physics},
  \emph{Springer Series on Atomic, Optical, and Plasma Physics}, vol.~68,
  Shevelko V, Tawara H (eds.). 2012; 125 -- 152.

\bibitem{litsarev-hci-2012}
Litsarev MS, Shevelko VP. Multiple-electron losses of highly charged ions
  colliding with neutral atoms. \emph{Physica Scripta}  2013;
  \textbf{2013}(T156):014\,037.

\bibitem{uspekhi2013}
Tolstikhina IY, Shevelko VP. Collision processes involving heavy many-electron
  ions interacting with neutral atoms. \emph{Physics-Uspekhi}  2013;
  \textbf{56}(3):213.

\bibitem{litsarev-jpb-2014}
Tolstikhina IY, Litsarev MS, Kato D, Song MY, J-S Y, Shevelko V. Collisions of
  be, fe, mo and w atoms and ions with hydrogen isotopes: electron capture and
  electron loss cross sections. \emph{Journal of Physics B: Atomic, Molecular
  and Optical Physics}  2014; \textbf{47}(3):035\,206.

\bibitem{litsarev-cpc-2013}
Litsarev MS. The deposit computer code: Calculations of electron-loss
  cross-sections for complex ions colliding with neutral atoms. \emph{Computer
  Physics Communications}  2013; \textbf{184}(2):432--439.

\bibitem{litsarev-cpc-2014}
Litsarev MS, Oseledets IV. The deposit computer code based on the low-rank
  approximations. \emph{Computer Physics Communications}  2014;
  \textbf{185}(10):2801–2802.

\bibitem{beylkin-expsum-2005}
Beylkin G, Monz{\'o}n L. On approximation of functions by exponential sums.
  \emph{Appl. Comput. Harm. Anal.}  2005; \textbf{19}(1):17--48,
  \doi{10.1016/j.acha.2005.01.003}.

\bibitem{beylkin-expsumrev-2010}
Beylkin G, Monz{\'o}n L. Approximation by exponential sums revisited.
  \emph{Appl. Comput. Harm. Anal.}  2010; \textbf{28}(2):131--149,
  \doi{10.1016/j.acha.2009.08.011}.

\bibitem{tee-mosaic-1996}
Tyrtyshnikov EE. Mosaic-skeleton approximations. \emph{Calcolo}  1996;
  \textbf{33}(1):47--57, \doi{10.1007/BF02575706}.

\bibitem{gtz-psa-1997}
Goreinov SA, Tyrtyshnikov EE, Zamarashkin NL. A theory of pseudo--skeleton
  approximations. \emph{Linear Algebra Appl.}  1997; \textbf{261}:1--21,
  \doi{10.1016/S0024-3795(96)00301-1}.

\bibitem{gtz-maxvol-1997}
Goreinov SA, Zamarashkin NL, Tyrtyshnikov EE. Pseudo--skeleton approximations
  by matrices of maximum volume. \emph{Mathematical Notes}  1997;
  \textbf{62}(4):515--519, \doi{10.1007/BF02358985}.

\bibitem{hackbra-expsum-2005}
Hackbusch W, Braess D. Approximation of {$\frac{1}{x}$} by exponential sums in
  {$[1,\infty]$}. \emph{IMA J. Numer. Anal.}  2005; \textbf{25}(4):685--697.

\bibitem{GHK-ten_inverse_ellipt-2005}
Gavrilyuk IP, Hackbusch W, Khoromskij BN. Tensor-product approximation to the
  inverse and related operators in high-dimensional elliptic problems.
  \emph{Computing}  2005; (74):131--157.

\bibitem{cross2dgit}
C++ code: Schur cross 2d ;
  \urlprefix\url{https://bitbucket.org/appl\_m729/schur\_cross2d}.

\bibitem{stoer2002}
Stoer J, Bulirsch R. \emph{Introduction to Numerical Analysis}. 3rd ed. edn.,
  Springer, 2002.

\bibitem{abramowitz-stegun}
Abramowitz M, Stegun IA. \emph{Handbook of Mathematical Functions with
  Formulas, Graphs, and Mathematical Tables}. tenth dover printing, tenth gpo
  printing edn., Dover: New York, 1972.

\bibitem{PhysRevA.36.467}
Salvat F, Martnez JD, Mayol R, Parellada J. Analytical
  dirac-hartree-fock-slater screening function for atoms (z=1-92). \emph{Phys.
  Rev. A}  1987; \textbf{36}:467--474, \doi{10.1103/PhysRevA.36.467}.

\bibitem{depositgit}
C++ code: Deposit 2014 ;
  \urlprefix\url{https://bitbucket.org/appl\_m729/code-deposit}.

\bibitem{slater1960}
Slater J. \emph{Quantum theory of atomic structure}. International series in
  pure and applied physics, McGraw-Hill, New York, 1960.

\bibitem{shevelko1993}
Shevelko VP, Vainshtein LA. \emph{Atomic physics for hot plasmas}. Institute of
  Physics Pub., 1993.

\bibitem{Khoromskij2006Kn}
Khoromskij B. Structured rank-(r1, . . . ,rd) decomposition of function-related
  tensors in rd. \emph{Comput. Methods Appl. Math.}  2006; \textbf{6}:194--220.

\end{thebibliography}

\end{document}